\documentclass[final,5p,times,preprint]{elsarticle}
\usepackage[utf8]{inputenc}
\usepackage{amsfonts}
\usepackage{amsmath}
\usepackage{amssymb}
\usepackage{bm}
\usepackage{lineno,hyperref}
\usepackage{comment}
\usepackage{xcolor}
\usepackage{booktabs}
\modulolinenumbers[5]
\usepackage[normalem]{ulem}

\journal{Physics Letters B}

\biboptions{comma,sort&compress}

\begin{document}

\begin{frontmatter}

\title{Linking structure and  dynamics in $(p,pn)$ reactions with Borromean nuclei: the $^{11}$Li$(p,pn){^{10}}$Li case}


\author[FAMN]{M. G\'omez-Ramos\corref{mail}}
\cortext[mail]{Corresponding author}
\ead{mgomez40@us.es}

\author[FAMN,ECT]{J. Casal}

\author[FAMN]{A. M. Moro}

\address[FAMN]{Departamento de F\'{\i}sica At\'omica, Molecular y Nuclear, Facultad de F\'{\i}sica, Universidad de Sevilla, Apartado 1065, E-41080 Sevilla, Spain}
\address[ECT]{European Centre for Theoretical Studies in Nuclear Physics and Related Areas (ECT$^*$) and Fondazione Bruno Kessler, Villa Tambosi, Strada delle Tabarelle 286, I-38123 Villazzano (TN), Italy}

\begin{abstract}
One-neutron removal $(p,pn)$ reactions induced by two-neutron Borromean nuclei are studied within a Transfer-to-the-Continuum (TC) reaction framework, which incorporates the three-body character of the incident nucleus. The relative energy distribution of the residual unbound two-body subsystem, which is assumed to retain information on the structure of the original three-body projectile, is computed by evaluating the transition amplitude for different neutron-core final states in the continuum. 
These transition amplitudes depend on the overlaps between the original three-body ground-state wave function and the two-body continuum states populated in the reaction, thus ensuring a consistent description of the incident and final nuclei. By comparing different $^{11}$Li three-body models, it is found that the $^{11}$Li$(p,pn){^{10}}$Li relative energy spectrum is very sensitive to the position of the $p_{1/2}$ and $s_{1/2}$ states in $^{10}$Li and to the partial wave content of these configurations within the $^{11}$Li ground-state wave function. The possible presence of a low-lying $d_{5/2}$ resonance is discussed. The coupling of the single particle configurations with the non-zero spin of the $^{9}$Li core, which produces a spin-spin splitting of the states, is also studied. Among the considered models, the best agreement with the available data is obtained with a $^{11}$Li model that incorporates the actual spin of the core and contains $\sim$31\% of $p_{1/2}$-wave content in the $n$-$^9$Li subsystem, in accord with our previous findings for the $^{11}$Li(p,d)$^{10}$Li transfer reaction, and a near-threshold virtual state.


\end{abstract}


\end{frontmatter}


\section{Introduction}\label{sec:intro}
Two-neutron Borromean nuclei are unique nuclear systems lying at the edge of the neutron drip-line. These are short-lived, weakly bound nuclei,  with typically no bound excited states, and whose binary subsystems are unbound.  Although some of them, such as $^{6}$He  and $^{11}$Li,  had already been identified long ago as products of reactions with stable beams \cite{Bje36,poskanzer66}, it was not until the late eighties that their unusual properties (such as their large size) were realized thanks to the pioneering experiments performed by Tanihata and collaborators \cite{Tanihata85} using secondary beams of these species and the subsequent theoretical works initiated by Hansen and Jonson \cite{Hansen87}. The picture emerging from these studies revealed a very exotic structure, consisting of a relatively compact core surrounded by two loosely bound nucleons forming a dilute halo.  


Later works have revealed that this fragile structure  arises from a delicate interplay of different effects, such as the pairing interaction between the halo neutrons or the coupling of the motion of these nucleons with tensor and collective excitations of the core (e.g.~\cite{Myo07,Barranco01,Ikeda10}). A quantitative account of all these effects is a challenging theoretical problem and, quite often, different models lead to different (sometimes contradictory) predictions of the structure properties, such as energies and spin-parity assignments. 

Experimentally, a successful technique to probe the properties of the neutron-core system is by means of $(p,pn)$ reactions at intermediate energies (above $\sim$100 MeV/nucleon), in which the radioactive beam collides with a proton target, removing one neutron, and leaving an (unbound) residual nucleus, which will eventually decay into a neutron and a core \cite{Aksyutina2008,Aksyutina2013}. Typically, these experiments measure the relative energy spectrum of this neutron-core system, whose prominent structures are associated with virtual states or resonances. Moreover, if the core is left in an excited state, gamma rays will be also emitted \cite{Kondo2010}. Angular momentum and spin assignment of these structures is often done by comparing these spectra with the profiles expected in the hypothetical neutron-core two-body scattering (e.g.~Breit-Wigner functions). This procedure is hampered by a number of limitations. For instance, it ignores completely the effect of the reaction dynamics on the spectra and, therefore, there is no a priori information on the absolute magnitude of the cross sections or, in other words, between the reaction observables and the underlying structure against which the data are confronted.  Moreover, due to energy resolution, resonances will appear smeared out or even unresolved. 


It is our purpose in this work to propose a new theoretical framework for the analysis of $(p,pn)$ reactions induced by Borromean nuclei in which  the structure model of the incident three-body nucleus is incorporated into a reaction formalism, thereby enabling the computation of reaction observables to be directly compared with the reaction data. In particular,  we will use a three-body model, which has been very successful in the understanding of the properties of Borromean nuclei (e.g.~\cite{Zhukov93}). For the reaction dynamics, we employ the Transfer-to-the-Continuum (TC) framework, which is formally similar to a CCBA (coupled-channel Born approximation) \citep{Asc69,Asc74} approach populating unbound states of the $p$-$n$ system. This method has already been applied to $(p,pn)$ reactions with two-body projectiles~\cite{Moro15}. We apply this formalism to describe the reaction $^{11}$Li(p,pn)$^{10}$Li at 280~MeV/A, which was measured a few years ago at GSI~\cite{Aksyutina2008}. 

The paper is organized as follows. In Sec.~\ref{sec:formal}, we present the reaction formalism, which is an extended version of that given in Ref.~\cite{Moro15} and will allow us to explain how the three-body structure enters into the calculation. In Sec.~\ref{sec:results}, the formalism is applied to the $^{11}$Li(p,pn)$^{10}$Li reaction, focusing on the relative-energy distribution of the decaying $^{10}$Li subsystem. Different structure models are considered and their impact on the reaction observables is discussed. Finally, in Sec.~\ref{sec:sum} we summarize the main results of the work and outline possible applications and extensions.

\section{Reaction Formalism}\label{sec:formal}

\begin{figure}[t]
\centering
\includegraphics[width=0.7\linewidth]{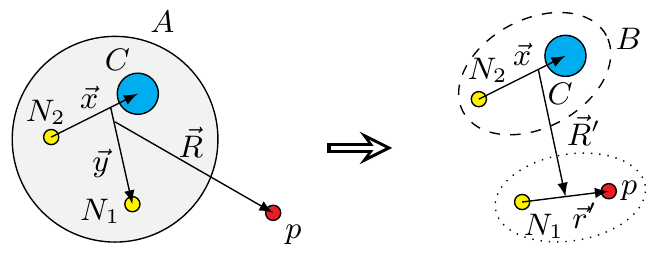}
\caption{Diagram for a $(p,pN)$ reaction induced by a three-body projectile in inverse kinematics.}
\label{fig:scheme}
\end{figure}

A $(p,pN)$ reaction induced by a three-body projectile comprising an inert core $(C)$ plus two valence neutrons $(N1, N2)$ takes the form, 
\begin{equation}
\underbrace{(C+N_1+N_2)}_A + p \rightarrow \underbrace{(C+N_2)}_{B} + N_1 + p,
\label{eq:scheme}
\end{equation}
which is schematically depicted in Fig.~\ref{fig:scheme}. If the nucleus $B$ does not form bound states (e.g., the composite $A$ is a Borromean system) the products of its decay after one neutron removal will provide spectroscopic information on the original projectile wave function. As in Ref.~\cite{plb1}, we describe the process using a {\it participant/spectator} approximation, assuming that the reaction occurs due to the interaction of the incident proton with a single neutron $(N_1)$ of $A$, whereas the subsystem $B=N_2+C$ remains unperturbed. The prior-form transition amplitude of such a process can be formally reduced to an effective few-body problem, leading to
\begin{equation}
\mathcal{T}_{if}=\sqrt{2}\langle \Psi_{f}^{(-)}(\vec{x},\vec{R}',\vec{r}')|V_{pN_1}+U_{pB}-U_{pA} |\Phi_A(\vec{x},\vec{y}) \chi_{pA}^{(+)}(\vec{R}) \rangle,
\label{eq:tmatrix}
\end{equation}
where $\Phi_A$ represents the ground-state wave function of the initial three-body composite, $\chi^{(+)}_{pA}$ is the distorted wave generated by the auxiliary potential $U_{pA}$, and $\Psi^{(-)}_f$ is the exact four-body wave function for the outgoing $pN_1$-$B$ system. The $\pm$ superscript refers to the usual ingoing or outgoing boundary conditions. Notice the explicit factor $\sqrt{2}$ arising from the two identical neutrons in the three-body projectile. The origin of this factor is further discussed in Ref.~\cite{GLENDENNING198345}.

To reduce Eq.~(\ref{eq:tmatrix}) to a tractable form, in the TC method we approximate the exact wave function $\Psi^{(-)}_f$ by the factorized expression,
\begin{equation}
\Psi_{f}^{(-)}(\vec{x},\vec{R}',\vec{r}') \approx \varphi^{(-)}_{\vec{q},\sigma_2,\zeta}(\vec{x}) \Upsilon_f^{(-)}(\vec{R}',\vec{r}')
\label{eq:tcwf}
\end{equation}
where $\varphi^{(-)}_{\vec{q},\sigma_2,\zeta}$ is a two-body continuum wave function with wave number $\vec{q}$ and definite spin projections of the binary subsystem $B$, and 
$\Upsilon_f^{(-)}$ is a three-body wave function describing the relative motion of the $p$-$N_1$-$B$ system in the exit channel. As in Ref.~\cite{Moro15}, we expand this function in $pN_1$ continuum states using a binning procedure~\cite{Austern87},
\begin{equation}
\Upsilon_f^{(-)}(\vec{R}',\vec{r}') \simeq \sum_{n\mathcal{J}\mathit{\Pi}}\phi_{n\mathcal{J}\mathit{\Pi}}(k_n,\vec{r}')\chi_{n\mathcal{J}\mathit{\Pi}}^{(-)}(K_n,\vec{R}').
\end{equation}
Here, $\phi_{n\mathcal{J}\mathit{\Pi}}$ are a set of $N$ discretized $pN_1$ bins with angular momentum $\mathcal{J}^\mathit{\Pi}$,
\begin{equation}
\phi_{n\mathcal{J}\mathit{\Pi}}(k_n,\vec(r)') = \sqrt{\frac{2}{\pi N}}\int_{k_{n-1}}^{k_n}\phi^{(+)}_{\mathcal{J}\mathit{\Pi}}(k,\vec{r}')dk,
\end{equation}
which are obtained from the scattering eigenstates $\phi^{(+)}_{\mathcal{J}\mathit{\Pi}}$ of the potential $V_{pN_1}$, and $\chi_{n\mathcal{J}\mathit{\Pi}}^{(-)}$ are distorted waves for the $pN_1$-$B$ relative motion. Note that the
subscript $f=\{n,\mathcal{J},\mathit{\Pi}\}$ in $\Upsilon_f^{(-)}$ retains the information on the definite final state.

The function  $\varphi^{(-)}_{\vec{q},\sigma_2,\zeta}$ in Eq.~(\ref{eq:tcwf}) is the time-reversed of $\varphi^{(+)}_{\vec{q},\sigma_2,\zeta}$, which can be written as~(c.f.~\cite{Satchler1983}, p.~135)
\begin{equation}
\begin{split}
\varphi^{(+)}_{\vec{q},\sigma_2,\zeta}(\vec{x}) &      = \frac{4\pi}{qx}\sum_{LJJ_TM_T}i^{L}Y^{*}_{LM}(\widehat{q}) 
						     \langle LMs_2\sigma_2|JM_J\rangle  \\
					    & \times \langle JM_JI\zeta|J_TM_T\rangle  f_{LJ}^{J_T}(qx) \left[\mathcal{Y}_{Ls_2J}(\widehat{x})\otimes\kappa_I\right]_{J_TM_T},
\end{split}
\label{eq:2bcont}
\end{equation}
where, for each component, the orbital angular momentum $\vec{L}$ and the spin $\vec{s}_2$ of the neutron $N_2$ couple to $\vec{J}$, and $\vec{J}_T$ results from coupling $\vec{J}$ with the spin $\vec{I}$ of the core. Note that, in our schematic notation, $\vec{x}$ contains also the internal coordinates of $C$. The radial functions $f_{LJ}^{J_T}$ are obtained by direct integration of the two-body Schrodinger equation for the $N_2+C$ system subject to standard scattering boundary conditions.

For the wave function of the projectile nucleus, $\Phi_A^{j\mu}(\vec{x},\vec{y})$, we use a three-body expansion in hyperspherical harmonics~\cite{Descouvemont03,FaCE,MRoGa05} using a pseudostate basis for the radial part called analytical transformed harmonic oscillator basis~\cite{JCasal13,JCasal14,JCasal15,JCasal16}. 

Assuming that the potentials $U_{pB}$ and $U_{pA}$ appearing in the transition amplitude (\ref{eq:tmatrix}) do not change the internal state of $B$, which is consistent with our participant/spectator approximation, one can perform the integral in the internal coordinates $\vec{x}$, giving rise to the overlap functions \cite{plb1}
\begin{equation}
\psi_{LJJ_TM_T}(q,\vec{y}) = \int \frac{f_{LJ}^{J_T}(qx)}{x}\left[\mathcal{Y}_{Ls_2J}(\widehat{x})\otimes\kappa_I\right]^*_{J_TM_T}\Phi_A^{j\mu}(\vec{x},\vec{y})d\vec{x},
\label{eq:overlap1}
\end{equation}
which contain all the relevant structure information. When used in Eq.~(\ref{eq:tmatrix}) one gets
\begin{equation}
\begin{split}
\mathcal{T}_{if}  & = \sqrt{2}\frac{4\pi}{q} \sum_{LJJ_TM_T} (-i)^{L} Y_{LM}(\widehat{q}) \langle LMs_2\sigma_2|JM_J\rangle \\ & \times\langle JM_JI\zeta|J_TM_T\rangle \mathcal{T}_{if}^{LJJ_TM_T},
\label{eq:tmatrix2}
\end{split}
\end{equation}
depending on a set of auxiliary CCBA-like amplitudes,
\begin{equation}
\mathcal{T}_{if}^{LJJ_TM_T} \equiv \langle \Upsilon_f^{(-)}|V_{pN_1}+U_{pB}-U_{pA} |\psi_{LJJ_TM_T}~\chi_{pA}^{(+)} \rangle,
\label{eq:tJ}
\end{equation}
These amplitudes enable a consistent description of the process, in which the three-body projectile and the binary fragment incorporate the same core-nucleon interaction. 

From the transition amplitude, and after integrating over the angles $\widehat{q}$ of the relative wave vector $\vec{q}$, the double differential cross section for a given final discretized bin $f=\{n,\mathcal{J},\mathit{\Pi}\}$ as a function of the $C$-$N_2$ relative energy and the scattering angle of $B$ with respect to the incident direction can be written as
\begin{align}
\frac{d\sigma^2_{n\mathcal{J}\mathit{\Pi}}}{d\Omega_B d\varepsilon_x} & = \frac{32\pi^2}{q^2}\rho(\varepsilon_x) \frac{1}{2(2j+1)}\frac{\mu_i\mu_f}{(2\pi\hbar^2)^2}\frac{k_n}{k_i} \nonumber \\
&\times\sum_{LJJ_T}\sum_{M_T\sigma_d}\left|\mathcal{T}_{if}^{LJJ_TM_T}\right|^2,
\label{eq:doubleTj}
\end{align}
where $\rho(\varepsilon_x)=\mu_x q/[(2\pi)^3 \hbar^2]$ is the density of $B$ states as a function of the $C$-$N_2$ excitation energy $\varepsilon_x$, $\mu_{i,f}$ the projectile-target reduced mass in the initial and final partitions, and $\sigma_d$ represents the spin projection of the $pN_1$ system. Although the non-relativistic expression for the cross section is shown in Eq.~(\ref{eq:doubleTj}) for simplicity, relativistic kinematics must be taken into account due to the high beam energy in typical $(p,pn)$ reactions \citep{Moro15}. Note that $q$ is related to the  $C$-$N_2$ relative energy as $q=\sqrt{2 \mu_x \varepsilon}/\hbar$, with $\mu_x$ its reduced mass. From Eq.~(\ref{eq:doubleTj}), the total differential cross section is obtained as an incoherent sum of the contributions to all $pN_1$ bins~\cite{Moro15}.

\section{Application to $^{11}$Li(p,pn)$^{10}$Li}\label{sec:results}

In this work, the formalism described in the preceding section is applied to the $^{11}$Li$(p,pn)^{10}$Li reaction, which has been measured  at GSI at 280~MeV/A, using inverse kinematics \citep{Aksyutina2008}. For the $V_{pn}$ potential, the Reid93 interaction \citep{Reid93} is chosen. The $p$-$^{11}$Li, $p$-$^{10}$Li and $n$-$^{10}$Li interactions are obtained as in Refs.~\cite{Moro15,plb1}, folding an effective $NN$ interaction with the ground-state density of the composite nucleus. For this purpose, the Paris-Hamburg $g$-matrix parametrization of the $NN$ interaction~\cite{Ger83,Rik84} is employed, while the ground-state densities are computed from Hartree-Fock calculations with the Sk20 effective interaction, using the code \textsc{OXBASH}~\cite{oxbash}. This folding is performed making use of the code \textsc{LEA}~\cite{lea}. Note that, due to the unbound nature of $^{10}$Li, the ground-state density of $^9$Li is used to generate the $N$-$^{10}$Li potentials.

As mentioned in the previous section, the $p$-$n$ continuum is discretized and truncated to a maximum angular momentum. The discretization is performed using a binning procedure~\cite{Austern87} with a step of $\Delta E=15$~MeV, although it was found that the studied observables are rather insensitive to the discretization used. In order to reduce the size of the calculations, we restrict the $p$-$n$ angular momentum to $\mathcal{J}_{max}=2$ in the exit channel and, as in Ref.~\cite{Moro15}, we ignore the couplings between different $\mathcal{J}^\mathit{\Pi}$ states. 
Test calculations for specific $n$-$^{9}$Li relative energies showed that these approximations do not modify significantly the shape of the energy distributions, although they underestimate the magnitude of the total cross section by about 10\%.
It must be remarked that no rescaling factors need to be applied to our calculations, since absolute cross sections can be obtained from the formalism. This allows us to assess the relative importance of different structure configurations to the cross section.

In the following, we explore the effects of the structure of $^{11}$Li on the $n$-$^9$Li relative energy spectrum after one neutron removal, performing the calculations in the low-energy range where the bulk of the cross section is concentrated. As in~\cite{plb1}, different potentials are used to generate the $\psi_{LJJ_TM_T}$ overlaps, leading to different structure properties of the $^{10}$Li continuum. In all calculations, we adopt the $^{11}$Li ground-state energy of $-0.37$ MeV \cite{Smith08}. We analyze the effect of virtual and resonant states on the computed spectra, studying in particular the effect of their splitting when the actual spin of the $^9$Li is included in the calculations. 

\subsection{Results ignoring  the $^9$Li spin}\label{subsec:spin0}

The spinless-core approximation has been widely used to describe the structure of $^{11}$Li \cite{Thompson94,Juanpi13,plb1}. The analysis of experiments involving $^{10,11}$Li usually assumes $I_{^9\mathrm{Li}}=0$ \cite{Blanchon2007,Aksyutina2008,Aksyutina2013,Cavallaro2017}, which simplifies the interpretation of the data. In this picture, the ground state of $^{10}$Li is a $s_{1/2}$ virtual state, followed by a low-energy $p_{1/2}$ resonance and, possibly, a $d_{5/2}$ state whose position and width is still unclear \cite{Blanchon2007,Cavallaro2017}. In this section we present three different models for $^{11}$Li which assume a spinless $^9$Li and allow us to study the influence of the structure properties on the reaction observables. Here, the $n$-$^9$Li interaction is modeled with central and spin-orbit Woods-Saxon terms adjusted to produce the $^{10}$Li virtual state and resonances at different positions, thus changing the partial wave content in the $^{11}$Li($0^+$) ground-state wave function. Some properties for $^{10}$Li and $^{11}$Li resulting from these potentials are shown in Table~\ref{tab:prop}. More details about these structure calculations can be found in Ref.~\cite{plb1}.




\begin{table*}[t]
\centering
\begin{tabular}{ccccccccccccc}
\toprule
   & \multicolumn{2}{c}{$a$ (fm)} & \multicolumn{2}{c}{$E_r[p_{1/2}]$ MeV} & $E_r[d_{5/2}]$ (MeV) & & \%$s_{1/2}$&\%$p_{1/2}$&\%$d_{5/2}$& $r_{mat}$ (fm) & $r_{ch}$ (fm)\\
\toprule
P3 & \multicolumn{2}{c}{-29.8} & \multicolumn{2}{c}{0.50} & 4.3 & & 64 & 30 & 3  & 3.6 & 2.48\\
P4 & \multicolumn{2}{c}{-16.2} & \multicolumn{2}{c}{0.23} & 4.3 & & 27 & 67 & 3  & 3.3 & 2.43\\
P5 & \multicolumn{2}{c}{-29.8} & \multicolumn{2}{c}{0.50} & 1.5 & & 39 & 35 & 23 & 3.2 & 2.42\\
\midrule
P1I &  -- & -37.9 & ~~0.37 & 0.61 & -- & & 67 & 31 & 1 & 3.2  & 2.41\\
\bottomrule
\end{tabular}
\caption{Features of the $^{10}$Li structure for the different potentials employed in this work. The second column shows the scattering length of the $s_{1/2}$ virtual state, the third column gives the energy of the $p_{1/2}$ resonance, and the fourth column corresponds to the position of the $d_{5/2}$ state. Note that for the model including the spin of the core, P1I, the $s_{1/2}, p_{1/2}$ configuration split into $1^-,2^-,1^+,2^+$. The splitting for the $d_{5/2}$ component is not considered, as this resonance is disregarded in the P1I potential. The next three columns show the partial wave content of $^{10}$Li configuration within the $^{11}$Li ground state, while the last two columns show its matter and charge radii.}
\label{tab:prop}
\end{table*}

\begin{figure}[t]
\centering
\includegraphics[width=0.9\linewidth]{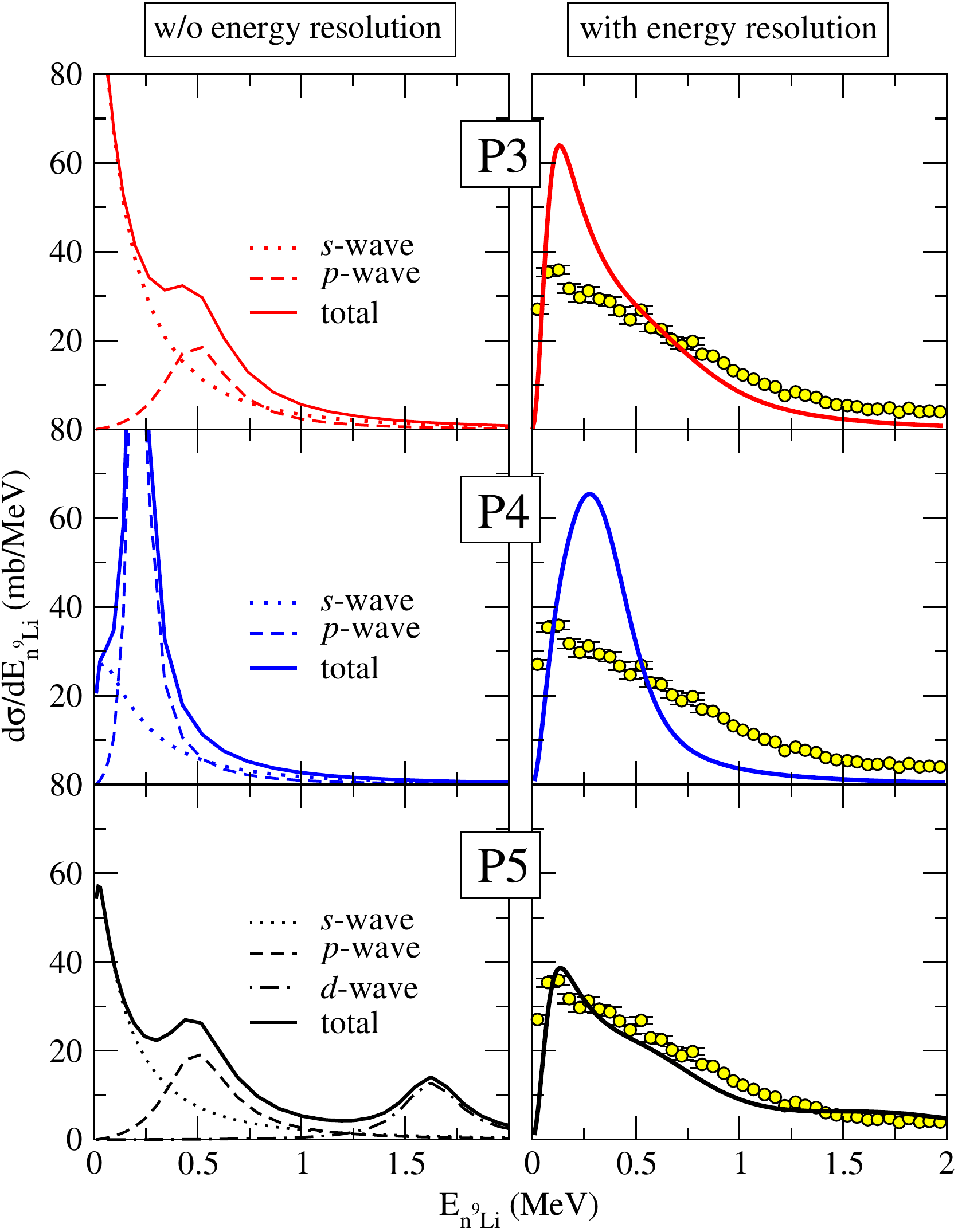}
\caption{Relative $n$-${^9\mathrm{Li}}$ energy spectrum for $^{11}$Li$(p,pn)^{10}$Li at 280 MeV/A. Calculations are presented for potentials P3, P4 and P5 with red, blue and black lines, respectively. In the left panels, the contributions for the $s$ and $p$ waves (and $d$ waves for P5) are shown along with their sum. In the right panels, the total cross section is shown after folding with the experimental resolution, along with experimental data from Ref.~\cite{Aksyutina2008}.}
\label{fig:nospin}
\end{figure}

The results of our calculation for models P3, P4 and P5 are presented  in the top, middle and bottom panels of Fig.~\ref{fig:nospin}, respectively. On the left side, we show the separate $s_{1/2}$ (dotted), $p_{1/2}$ (dashed) and, in the case of P5, $d_{5/2}$ (dot-dashed) contributions, together with their  sum (solid).
In models P3 and P4, the $d$-wave content in the $^{11}$Li ground state is very small, and the $d_{5/2}$ resonance appears at high energies (see Table~\ref{tab:prop}), thus making this contribution negligible. Note that the 
height of the peak of the $s_{1/2}$ contribution is related not only to the scattering length, but also to the its relative weight in the $^{11}$Li ground state. On the right side of Fig.~\ref{fig:nospin}, the total cross section is shown for the three models after the convolution with the experimental resolution and compared with the data from~\cite{Aksyutina2008}. It can be seen that, for P3, the main contribution to the cross section comes from $s_{1/2}$ states, while for P4 it comes from $p_{1/2}$ states. We see that both calculations fail to reproduce the shape of the experimental data, which is heavily influenced by the experimental resolution. In general, both P3 and P4 calculations seem to give too much cross section at low energies while too little at higher energies. The model P5, on the contrary, gives a better agreement with the experimental data by just lowering the position of the $d_{5/2}$ resonance. A similar effect was found in a previous work describing the knockout reaction on carbon~\cite{Blanchon2007}. Notice that models P3 and P5 provide the same $^{10}$Li states for the $s_{1/2}$ and $p_{1/2}$ configurations, but the partial wave content of $^{11}$Li is strongly affected by the presence of the $d_{5/2}$ $^{10}$Li resonance at low energies.

However, the $d$-wave content given by P5, 23\%, is rather large compared to the most recent experimental study~\cite{Aksyutina2013}, that amounts to $\sim$10\%. Moreover, the $d$ resonance has been recently identified at higher excitation energies in a $(d,p)$ experiment~\cite{Cavallaro2017}.
This, together with the oversimplification of the $^{10,11}$Li models neglecting the spin of the $^{9}$Li core, may indicate that the good agreement found for P5 is biased by the low energy resolution of the data. 



\subsection{Results including the $^9$Li spin}\label{subsec:spin}

By considering explicitly the spin of the $^9$Li core, $I=3/2^-$, the $n$-${^9\mathrm{Li}}$ single-particle configurations $s_{1/2}$ and $p_{1/2}$ split in $1^-,2^-$ and $1^+,2^+$ states, respectively. Since the $s$- and $p$-wave contributions dominate the low-energy relative energy spectra presented in the previous section, these doublets can affect the shape of the distributions. Previous theoretical studies including this effect have reported the existence of one or two $s$ virtual states, and two $p$ resonances \cite{Nun96a,Garrido02,Kik13}, although no direct experimental evidence of this splitting has been reported \cite{Fortune16}. In this section, we perform the same calculation as before, but using a structure model which couples the spin of $^9$Li to the angular momenta of the two neutrons. We have chosen the potential P1I from Ref.~\cite{plb1}, which we found to give a good description of $^{11}$Li$(p,d)^{10}$Li data~\cite{Sanetullaev16}. Some structure features obtained with this model are also shown in Table~\ref{tab:prop}.

\begin{figure}[t]
\centering
\includegraphics[width=0.85\linewidth]{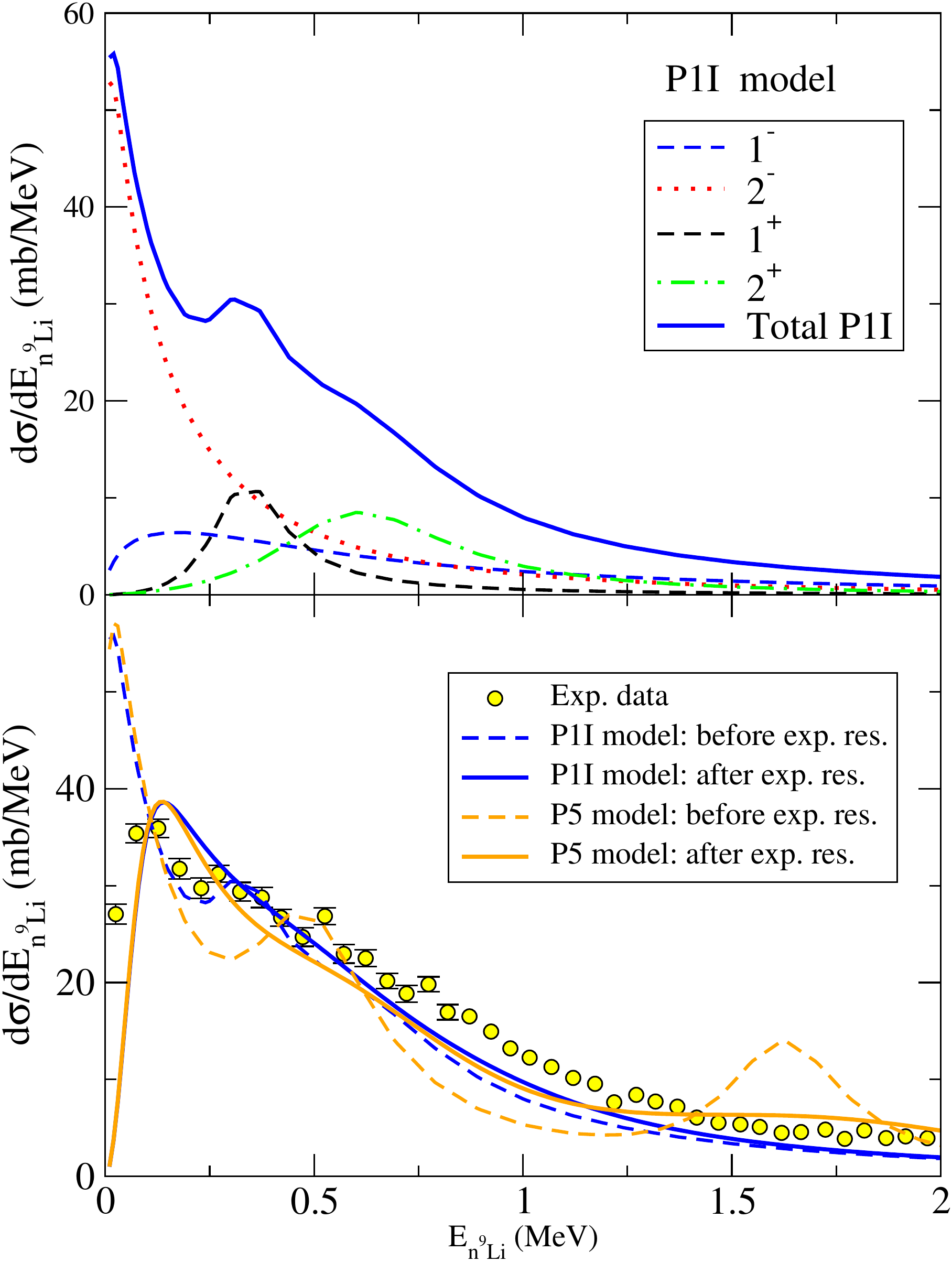}
\caption{Effect of the spin-spin splitting in the relative $n$-${^9\mathrm{Li}}$ energy spectrum. In the top panel, the different contributions for the $s$ ($1^-,2^-$) and $p$ ($1^+,2^+$) waves within the model P1I are presented. The total cross section is given by the solid blue line.  In the bottom panel, the results for models P1I and P5 are shown before and after folding with the experimental resolution. For P5, this includes also the $d_{5/2}$ states. The experimental data are from Ref.~\cite{Aksyutina2008}.}
\label{fig:spin32}
\end{figure}

In Fig.~\ref{fig:spin32}, results for P1I are presented along with those for P5. In the top panel, the contributions from the different two-body configurations, namely, $1^-$ (thin solid), $2^-$ (dotted), $1^+$ (dashed) and $2^+$ (dot-dashed) are presented, together with their sum  (thick solid). In this model, the 2$^-$ ground state of $^{10}$Li is characterized by a scattering length of $a_s=-37.9$ fm, and the $1^-$ states correspond to non-resonant continuum. The two $p$ resonances are obtained at 0.37 and 0.61 MeV, providing a doublet that could not be resolved experimentally. As shown in Table~\ref{tab:prop}, the three-body ground-state wave function probabilities in the $n$-$^9$Li subsystem are given by 31\% of $p_{1/2}$ components, 67\% of $s_{1/2}$ components and a negligible $d_{5/2}$ contribution. The weights of the individual $1^-,2^-,1^+,2^+$ configurations are 27\%, 40\%, 12\% and 19\%, respectively. Taking these values into consideration, the effective scattering length of the 2$^-$ ground state is reduced to $a_\text{eff}=-29.3$ fm, and the two $p$ resonances have their centroid at 0.52 MeV. 
In the bottom panel, the total cross section is presented before and after convoluting with the experimental resolution and compared with the experimental data. Blue lines correspond to model P1I including the spin-spin splitting, while orange lines correspond to the spinless-core model P5. We find the calculation using P1I to provide an even better agreement with the data, when compared to P5. This agreement seems to stem from two sources: first, the $p_{1/2}$ resonance is split, leading to a broader distribution that accommodates the high-energy tail shown by experimental data. Second, the splitting of the $s_{1/2}$ virtual state at low energies enables a reduction of the cross section in this region. Note that, in the present work, these doublets are obtained by introducing a core-spin dependence in the $n$-$^9$Li potential through a spin-spin term. This allows us to describe the splitting schematically, although the actual mechanism might involve more complex correlations, such as pairing \cite{Orrigo2009}, tensor correlations \cite{Myo07} or coupling to excited states of the core \cite{Nun96a}. 

In contrast to P5, the model P1I does not require a $d_{5/2}$ resonance at 1.5~MeV to reproduce the experimental data. The introduction of such a resonance at higher energies (as in models P3, P4 in the preceding section), would have little effect on the partial wave content of $^{11}$Li, thus preserving the agreement with the data. However, the coupling of the single-particle configuration $d_{5/2}$ to the $3/2^-$ of the core leads to $1^-,2^-,3^-$ and $4^-$ states from which no information exists, thus complicating the theoretical description of the possible resonances and the interpretation of the data. Clearly, this situation calls for more elaborate theoretical studies and experiments with better energy resolution.

\subsection{Factorization of the cross section}\label{sec:factorization}

\begin{figure}[ht]
\centering
\includegraphics[width=0.85\linewidth]{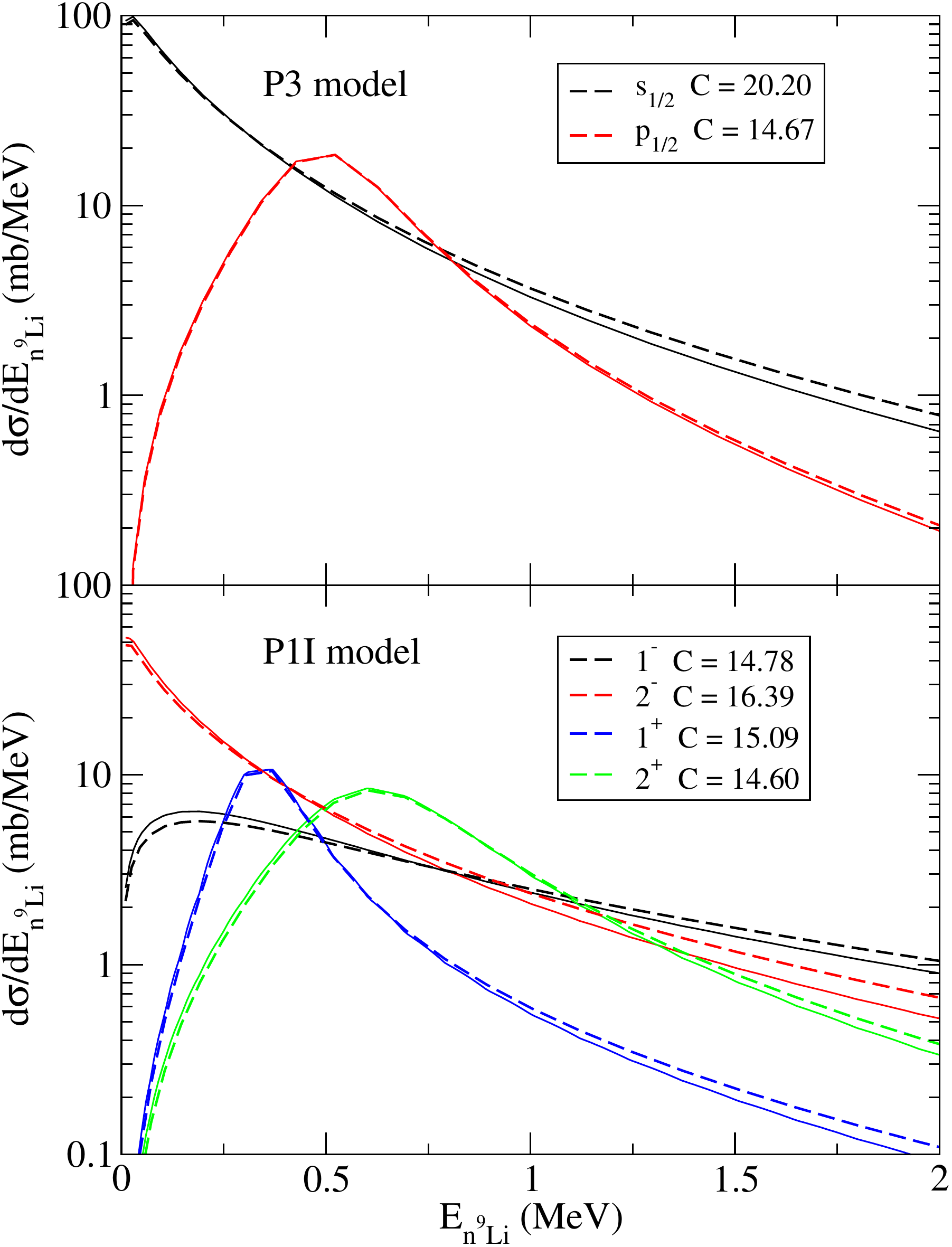}
\caption{Comparison between the TC calculations (solid lines) and the rescaled structure overlaps (dashed lines) for models P3 (top panel) and P1I (bottom panel) in logarithmic scale. The factor $\mathcal{C}^{LJJ_T}$ for each contribution is given in the legend. Calculations are shown without convoluting with the experimental resolution.}
\label{fig:sfcomp}
\end{figure}

The analysis of the experimental data presented in Ref.~\cite{Aksyutina2008} uses a factorization approximation for the cross section. The relative $n$-${^9\mathrm{Li}}$ energy spectrum is fitted with two distributions corresponding to the $s_{1/2}$ and $p_{1/2}$ states, with their relative weights as parameters. In addition to ignoring the partial wave content of the initial $^{11}$Li projectile, which modulates the relevance of the different components, this implies the assumption that the reaction dynamics introduce only a global scaling factor over some structure form factors. Under such considerations, the differential cross section for a given $(LJ)J_T$ configuration can be schematically written as
\begin{equation}
\frac{d\sigma^{LJJ_T}}{d\varepsilon_x} \simeq \mathcal{C}^{LJJ_T} K(\varepsilon_x) \eta^{LJJ_T}(\varepsilon_x),
\label{eq:sf}
\end{equation}
where $\eta^{LJJ_T}$ represents the structure form factor, $K$ is a kinematic function which contains the density of states and and all relevant constants, and $\mathcal{C}^{LJJ_T}$ is a global scaling factor, which contains the effect of the reaction mechanism. In our approach, $\eta^{LJJ_T}$ are given by the square of the structure overlaps as a function of the energy. This allows us to compare the shape of the cross sections in Eq.~(\ref{eq:sf}) with the results from our full TC calculations and, if possible, extract the $\mathcal{C}^{LJJ_T}$ factors. This is shown in Fig.~\ref{fig:sfcomp} for models P3 and P1I. The rescaled overlaps are rather similar to the full TC calculations, with only small deviations in the shape. This deviation is more significant for the $s$-wave contributions, which can be associated with the extended halo wave function of $^{11}$Li playing a role for the reaction dynamics, but it is still a minor effect. The resulting $\mathcal{C}^{LJJ_T}$ factors are similar in both models for the $p$-wave components and differ significantly for the $s$-wave. For models P4 and P5, the the $s$- and $p$-wave scaling factors are found to be almost identical to those for P3, while the $d$-wave scaling in the model P5 is similar to that for $p$-waves. However, even though at first order the reaction dynamics introduces only $J_T$-dependent factors, a proper reaction formalism is required to obtain them unambiguously. This, together with the role played by the weights of the different configurations in the ground state of the projectile, indicates that only a consistent description of both the structure and dynamics can provide a reliable interpretation of the data. 

\section{Summary and conclusions}\label{sec:sum}
We have presented a new method to study $(p,pn)$ reactions induced by three-body Borromean nuclei. The formalism is a natural extension of the  Transfer-to-the-Continuum method \cite{Moro15}, recently proposed and applied to two-body projectiles, to the case of three-body projectiles.  The model assumes a participant/spectator picture, in which the proton target knocks out one of the halo neutrons (the participant), while leaving unperturbed the remaining neutron-core subsystem (the spectator).  A key feature of the model is the use of structure overlaps obtained from a three-body model of the ground-state wave-function of the Borromean nucleus and the two-body scattering states of the neutron-core residual system.  
These overlaps are used in a CCBA-like prior-form transition amplitude, thus providing a connection between the structure model and the reaction observables without the need of introducing arbitrary scaling factors. 
 In particular, the formalism provides double differential cross sections as a function of the scattering angle of the residual two-body system and the relative energy between its constituents.  
 
The model has been applied to the $^{11}$Li$(p,pn){^{10}}$Li reaction at 280~MeV/A, comparing with  available data for the neutron-$^{9}$Li system relative-energy distribution.  Several structure models of $^{11}$Li have been compared, differing on the position of the assumed $s_{1/2}$ virtual states and $p_{1/2}$ resonances, and on the inclusion or not of the $^{9}$Li spin which, in turn, give rise to different relative weights for these partial waves in the $^{11}$Li(g.s.).   The calculated reaction observable is  found to be very sensitive to these structure properties. Among the considered models, the best agreement with the data is obtained using a $^{11}$Li model that incorporates the actual spin of the core and contains $\sim$31\% of $p_{1/2}$-wave content in the $n$-$^9$Li subsystem and a near-threshold virtual state with an effective scattering length of about -29~fm. The agreement stems from the splitting of the $s_{1/2}$ virtual state and the $p_{1/2}$ resonance. This splitting was obtained thanks to a spin-spin interaction in this work, although its actual origin may arise from more complex correlations. Interestingly, this model was found to provide also a good description of the recent $^{11}$Li($p$,$d$)$^{10}$Li transfer data measured at TRIUMF.

We have discussed also the possible presence of a low-lying $d$-wave resonance in $^{10}$Li. An overall good agreement with the data can be obtained in the model ignoring the $^9$Li spin by forcing a $d_{5/2}$ resonance to appear at $E_r$=1.5~MeV, which also reduces the $s$-wave content in $^{11}$Li. However, in view of other experimental evidences, the agreement might be merely accidental. In fact, such a resonance is not required in the model including the spin of $^9$Li to achieve a good description of the data. Due to the smearing effect produced by the energy resolution of the experiment, it is clear that further data, more sensitive to higher excitation energies and with better energy resolution, will certainly help in extracting robust conclusions on the $d_{5/2}$ states.

The formalism presented could be applied to study $(p,pn)$ or $(p,2p)$ reactions induced by other Borromean nuclei. Calculations of this kind for $^{14}$Be, including also the effect of core excitations, are in progress and will be presented elsewhere.



\section*{Acknowledgements}
This work has received funding from the Spanish Ministerio de Economía y Competitividad under Project No.~FIS2014-53448-C2-1-P and by the European Union Horizon 2020 research and innovation program under Grant Agreement No.~654002. M.G.-R.~acknowledges support from the Spanish Ministerio de Educaci\'on, Cultura y Deporte, Research Grant No.~FPU13/04109.

\section*{References}

\bibliography{mybibfile}

\end{document}